# Short-Pulse, Compressed Ion Beams at the Neutralized Drift Compression Experiment

P A Seidl[1], J J Barnard[2], R C Davidson[3], A Friedman[2], E P Gilson[3], D Grote[2], Q. Ji[1], I D Kaganovich[3], A Persaud[1], W L Waldron[1] and T Schenkel[1]

[1]Lawrence Berkeley National Laboratory, Berkeley, California, USA
[2]Lawrence Livermore National Laboratory, Livermore, California, USA
[3] Princeton Plasma Physics Laboratory, Princeton, New Jersey, USA

PASeidl@lbl.gov

**Abstract.** We have commenced experiments with intense short pulses of ion beams on the Neutralized Drift Compression Experiment (NDCX-II) at Lawrence Berkeley National Laboratory, with 1-mm beam spot size within 2.5 ns full-width at half maximum. The ion kinetic energy is 1.2 MeV. To enable the short pulse duration and mm-scale focal spot radius, the beam is neutralized in a 1.5-meter-long drift compression section following the last accelerator cell. A short-focal-length solenoid focuses the beam in the presence of the volumetric plasma that is near the target. In the accelerator, the line-charge density increases due to the velocity ramp imparted on the beam bunch. The scientific topics to be explored are warm dense matter, the dynamics of radiation damage in materials, and intense beam and beam-plasma physics including select topics of relevance to the development of heavy-ion drivers for inertial fusion energy. Below the transition to melting, the short beam pulses offer an opportunity to study the multi-scale dynamics of radiation-induced damage in materials with pump-probe experiments, and to stabilize novel metastable phases of materials when short-pulse heating is followed by rapid quenching. First experiments used a lithium ion source; a new plasma-based helium ion source shows much greater charge delivered to the target.

## 1. Introduction

Intense pulses of ions in the MeV range enable new studies of the properties of matter ranging from low intensity (negligible heating, but active collective effects due to proximate ion trajectories in time and space), to high intensity where the target may be heated to the few-eV range and beyond. By choosing the ion mass and kinetic energy to be near the Bragg peak, dE/dx is maximized and a thin target may be heated with high uniformity [1], thus enabling high-energy density physics (HEDP) experiments in the warm dense matter (WDM) regime. The Neutralized Drift Compression Experiment (NDCX-II) was designed with this motivation [2-4].

Reproducible ion pulses (N>$10^{11}$ /bunch), with bunch duration and spot size in the nanosecond and millimeter range, meet the requirements to explore the physics topics identified above. The formation of the bunches generally involves an accelerator beam with high perveance and low emittance, attractive for exploring basic beam physics of general interest, and relevant to the high-current, high-intensity ion beams needed for heavy-ion-driven inertial fusion energy [5].

Furthermore, short ion pulses at high intensity (but below melting) enable pump-probe experiments that explore the dynamics of radiation-induced defects in materials [6]. For high peak currents and short ion pulses, the response of the material to radiation may enter a non-linear regime due to the overlapping collision cascades initiated by the incident ions. These effects may be transient (no memory effect at a subsequent pulse) and the short, intense pulses of ions provide an opportunity to observe the time-resolved multi-scale dynamics of radiation-induced defects [6-8]. In addition, by measuring the ion range during the course of the ion pulse, the effects of defects and heating on range can be observed. The time-resolved information provides insight and constraints on models of defect formation and in the design of structural materials, for example, for fission and fusion reactors.

## 2. New helium ion source and opportunities for intense ion beam physics

The first target experiments used beam pulses of $Li^+$ accelerated to 1.2 MeV and focused to a beam radius, r ≅ 1mm and duration of 2 ns FWHM (peak current ~18 $A/cm^2$). These conditions were used to first commission the integrated accelerator components and then in target experiments, for example, to characterize dose rate effects on the ionoluminescence of yttrium aluminium perovskite (YAP). Single-shot YAP scintillator streak spectrometer results showed wavelength structures and motivate follow-up measurements with other materials while varying the focused intensity. These and other results are described in Ref. [9].

Recently, we have installed and used a new multicusp, multiple-aperture plasma ion source. The source can generate high purity ion beams of, for example, protons, helium, neon and argon. To date, we have used it solely for the generation of $He^+$ ions, where the source injects significantly greater charge than the lithium ion source. Furthermore, helium at about 1 MeV is nearly ideal for highly uniform volumetric energy deposition, because particles enter thin targets slightly above the Bragg peak energy and exit below it, leading to energy loss in the target, uniform within several percent.

The large extraction area is a novel feature of this new plasma source (38 $cm^2$). To control the plasma meniscus geometry over such a large area and maintain low emittance, the ion extraction gap is established between two parallel plates with many aligned holes, millimeter-diameter with a few-millimeter pitch. A filament driven plasma is formed for 1 millisecond. The plasma facing hole plate, which defines the ion-emission surface, is biased to nearly the same potential as the filament. One kilovolt ions are extracted between the parallel hole plates during the 1-μs high voltage pulse applied via voltage division to both the 3-mm gap and the ≈0.5 m injector column.

The average extracted current density over the 7-cm diameter emission plane is in the range 1-5 $mA/cm^2$, depending on the voltage, helium gas flow, and filament discharge settings, which are controlled stably and with high reproducibility. The beam is formed with low emittance ($\varepsilon_n$ < 2 mm•mrad estimated from simulations) and with a manageable gas load into the downstream accelerating structure. Particle-in-cell simulations helped determine injector voltage settings so that the ion trajectories are similar to those of the previous lithium ion source in the accelerator. Operating parameters of the new ion source are summarized in Table 1, and a fuller description of the source design and characterization may be found in Ji et al. [10].

Table 1. Operating parameter ranges for the new helium ion source in NDCX-II.

| | | | |
|---|---|---|---|
| $I_{arc}$ | 1-4 A | Flow rate | 45 standard cc/min |
| $V_{filament\ bias}$ | 100 V | $P_{source}$ | 3.5 x $10^{-3}$ Torr |
| $P_{filament}$ | 5A x 50 V | $P_{injector}$ | 8 x $10^{-6}$ Torr |
| $V_{injector}$ | 135 kV | $P_{accelerator}$ | ≤3 x $10^{-6}$ Torr |

An ion induction accelerator is capable of simultaneously accelerating and rapidly compressing beam pulses by adjusting the slope and amplitude of the voltage waveforms in each gap. In NDCX-II, this is accomplished with 12 compression and acceleration waveforms driven with peak voltages ranging from 15 kV to 200 kV and durations of 0.07-1 μs.

The first seven acceleration cells are driven by spark-gap switched, lumped element circuits tuned to produce the required cell voltage waveforms. These waveforms ("compression" waveforms because of their characteristic triangular shape) have peak voltages ranging from 20 kV to 50 kV. An essential design objective of the compression pulsers is to compress the bunch to <70 ns so that it can be further accelerated and bunched by the 200-kV Blumlein pulsers which drive the last five acceleration cells.

In the final drift section, the bunch has a head-to-tail velocity ramp that further compresses the beam by an order-of-magnitude. The space-charge forces are sufficiently high at this stage to require that dense plasma, generated prior to the passage of the beam pulse, neutralizes the beam self-field and to enables focusing and bunching of the beam to the millimeter and nanosecond range [11]. Since the coasting beam compresses with negligible space-charge repulsion, the spot size and duration are limited mainly by voltage waveform fidelity and the chromatic aberrations and emittance of the beam.

The accelerator provides a novel platform to extend the limits of intense beam and beam-plasma physics. For example, the ion-electron two-stream instability has been predicted [12], and it may be observed in NDCX-II by passing a nearly constant energy beam through a plasma. The manifestation would be a significant transverse defocusing and longitudinal bunching of the specially prepared beam. In some parameter regimes in an inertial fusion energy reactor chamber, this could be a critical issue for the transport of a much higher current ion beam in the presence of a significant number of background electrons. In NDCX-II, the effect is normally absent because of the imposed velocity ramp on the beam distribution.

Another interesting opportunity would be to demonstrate the collective focusing of an ion beam in a weak magnetic field [13]. The focusing occurs due to the formation of a strong radial self-electric field due to the rearrangement of the plasma electrons moving with the beam, and in response to a weak magnetic field. The magnetic field is established by the final solenoid near the end of the neutralized drift section. But, instead of using the full strength of the final solenoid (8 Tesla) to focus the beam, equivalent focusing may be achieve with only ~0.02 Tesla. One challenge is to limit the source of electrons to the plasma region upstream of the solenoid. If demonstrated to be practical, the focusing magnet strength would be greatly reduced with a corresponding reduction of the stray magnetic field at the nearby target plane – as preferred for some beam-target interaction experiments.

Non-neutral drift compression and beam pulse manipulation is of interest for future high intensity machines. For example, a nearly monoenergetic beam may be formed at the axial location where the significant velocity tilt is removed by the longitudinal space charge of the intense beam pulse, attractive for reducing chromatic effects when focusing to small spot size. The flexibility to independently trigger and shape the waveforms of each induction cell allows the formation of varied pulse shapes, such as foot—peak waveforms, as generally desired for indirectly driven fusion targets. Finally, double pulses of $K^+$ or $Li^+$ ions, separated by 0.05 - 0.5 μs, have been created from single triggers of the high voltage pulses in the injector and accelerator [6]. A single bunch is injected into the accelerator but was cleaved due to the shifted timings of the firing of the compression waveforms. Double pulses are of interest for example, in pump-probe experiments of radiation effects in matter.

## 3. Recent results and outlook

Using the new helium ion source, the transverse distribution of the focused beam at the target plane is characterized by a thin scintillator and a gated image-intensified CCD camera. As shown in Fig. 1, the beam distribution is peaked with 80% of the charge within a radius of 0.8 mm.

Previously, for the $Li^+$ beam, the charge in the highly compressed bunch was 1-2 nC. Simulations indicate that the focused charge is expected to ultimately reach over 30 nC per pulse while maintaining a small focal spot size and bunch duration, since much greater charge is now available from the new helium source. Toward that goal, we have measured 15 nC ($9x10^{10}$ ions/pulse) in a few-mm focal spot with the $He^+$ beam (Fig. 2). A near-term objective is to achieve a pulse duration of 1-2 ns by model-derived tuning of the acceleration and bunching waveforms and the solenoid field strengths.

With the range of intensities and dose rates presently available, there is an opportunity to create novel states of matter with the NDCX-II beam, such as local nitrogen-vacancy center formation in nitrogen-implanted diamond by transient heating, followed by rapid quenching [14]. Annealing of specially prepared materials often takes place on a timescale of seconds to minutes, such as in the formation of paramagnetic materials [15]. With NDCX-II, we have an opportunity to locally excite materials with intense ion pulses on a ns timescale and stabilize novel phases by rapid quenching.

Intense ion beams from induction accelerators have complementary advantages (e.g., low energy spread, benign radiation environment) vs. beams derived from laser-plasma acceleration [16].

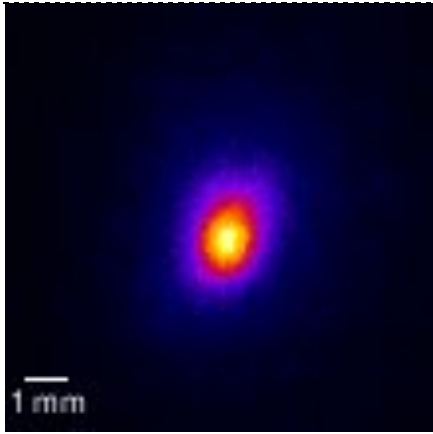

**Figure 1.** The transverse (x, y) distribution of the He$^+$ beam is measured with a scintillator and CCD camera. With the new helium source, NDCX-II can provide 1-MeV helium ion beams with an energy deposition greater than 0.7 joules/cm$^2$ (~40 A/cm$^2$).

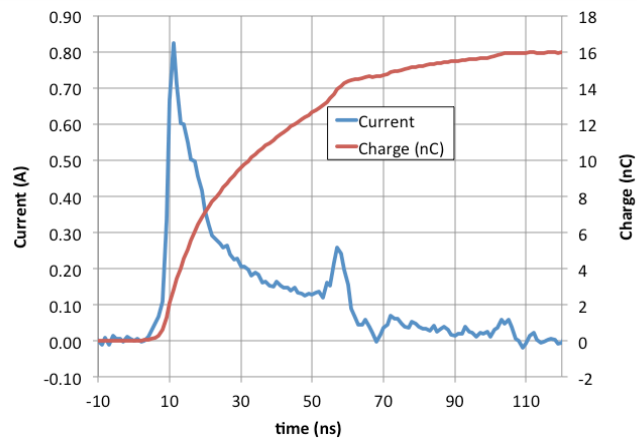

**Figure 2.** The full temporal distribution of the pulses is ~50 ns with a ~1 ns rise time, though still with a broader tail than desired. The integrated charge in the pulse is ≈15 nC. The pulse was measured with a fast Faraday collector isolated from the neutralizing plasma with hole plates with low geometric transparency.

**Acknowledgments**
This work was supported by the Office of Science of the US Department of Energy under contracts DE-AC02–05CH11231 (LBNL), DE-AC52-07NA27344 (LLNL) and DE-AC02-09CH11466 (PPPL).